\documentclass{ws-ijmpa}

\def\buildrel#1\over#2{\mathrel{
 \mathop{\kern 0pt#2}\limits^{#1}}}
\def\bbuildrel#1_#2^#3{\mathrel{
 \mathop{\kern 0pt#1}\limits_{#2}^{#3}}}

\def\Diga{\kern 0.25em\
\put(20,0){\circle*{2}} 
\qbezier(0, 0)(10, 15)(20, 0)
\qbezier(0, 0)(10, -15)(20, 0)\kern 1.9em\ 
}

\def\Digb{\kern 0.25em\
\put(0,0){\circle*{2}}
\put(20,0){\circle*{2}}
\qbezier(0, 0)(10, 15)(20, 0)
\qbezier(0, 0)(10, -15)(20, 0)\kern 1.9em\  
}

\def\Digc{\kern 0.25em\
\put(10, 7.2){\circle*{2}}
\put(0,0){\circle*{2}}
\put(20,0){\circle*{2}}
\qbezier(0, 0)(10, 15)(20, 0)
\qbezier(0, 0)(10, -15)(20, 0)\kern 1.9em\  
}

\def\Digd{\kern 0.25em\
\put(10, -7.5){\circle*{2}}
\put(10, 7){\circle*{2}}
\put(0,0){\circle*{2}}
\put(20,0){\circle*{2}}
\put(0,0){\line(20,0){20}}
\qbezier(0, 0)(10, 15)(20, 0)
\qbezier(0, 0)(10, -15)(20, 0)\kern 1.9em\  
}

\def\Dige{\kern 0.25em\
\put(6, -6.5){\circle*{2}}
\put(14, -6.5){\circle*{2}}
\put(0,0){\circle*{2}}
\put(20,0){\circle*{2}}
\put(0,0){\line(20,0){20}}
\qbezier(0, 0)(10, 15)(20, 0)
\qbezier(0, 0)(10, -15)(20, 0)\kern 1.9em\ 
}

\def\Digf{\kern 0.25em\
\put(10, -7.5){\circle*{2}}
\put(0,0){\circle*{2}}
\put(20,0){\circle*{2}}
\put(0,0){\line(20,0){20}}
\qbezier(0, 0)(10, 15)(20, 0)
\qbezier(0, 0)(10, -15)(20, 0)\kern 1.9em\  
}

\def\Digg{\kern 0.25em\
\put(0,0){\circle*{2}}
\put(20,0){\circle*{2}}
\put(0,0){\line(20,0){20}}
\qbezier(0, 0)(10, 15)(20, 0)
\qbezier(0, 0)(10, -15)(20, 0)\kern 1.9em\  
}

\def\Digh{\kern 0.25em\
\put(10, -7.5){\circle*{2}}
\put(10, 0){\circle*{2}}
\put(0,0){\circle*{2}}
\put(20,0){\circle*{2}}
\put(0,0){\line(20,0){20}}
\qbezier(0, 0)(10, 15)(20, 0)
\qbezier(0, 0)(10, -15)(20, 0)\kern 1.9em\  
}

\def\Figa{\kern 0.25em\
\put(20,0){\circle*{2}}
\qbezier[30](0, 0)(10, 15)(20, 0)
\qbezier[30](0, 0)(10, -15)(20, 0)\kern 1.9em\  
}

\def\Figb{\kern 0.25em\
\put(0,0){\circle*{2}}
\put(20,0){\circle*{2}}
\qbezier[30](0, 0)(10, 15)(20, 0)
\qbezier[30](0, 0)(10, -15)(20, 0)\kern 1.9em\  
}

\def\Figc{\kern 0.25em\
\put(6, 0){\circle*{2}}
\put(14, 0){\circle*{2}} 
\put(0,0){\circle*{2}}
\put(20,0){\circle*{2}}
\put(0,0){\line(20,0){20}}
\qbezier[25](0, 0)(10, 15)(20, 0)
\qbezier[25](0, 0)(10, -15)(20, 0)\kern 1.9em\ 
}

\def\Figd{\kern 0.25em\
\put(10, -8){\circle*{2}}
\put(0,0){\circle*{2}}
\put(20,0){\circle*{2}}
\put(0,0){\line(20,0){20}}
\qbezier[25](0, 0)(10, 15)(20, 0)
\qbezier[25](0, 0)(10, -15)(20, 0)\kern 1.9em\  
}

\def\Fige{\kern 0.25em\
\put(10, 0){\circle*{2}}
\put(0,0){\circle*{2}}
\put(20,0){\circle*{2}}
\put(0,0){\line(20,0){20}}
\qbezier[25](0, 0)(10, 15)(20, 0)
\qbezier[25](0, 0)(10, -15)(20, 0)\kern 1.9em\ 
}

\def\Figf{\kern 0.25em\
\put(10, -7.5){\circle*{2}}
\put(10, 7.5){\circle*{2}}
\put(0,0){\circle*{2}}
\put(20,0){\circle*{2}}
\put(0,0){\line(20,0){20}}
\qbezier[25](0, 0)(10, 15)(20, 0)
\qbezier[25](0, 0)(10, -15)(20, 0)\kern 1.9em\ 
}

\def\Figg{\kern 0.25em\
\put(10, 7.5){\circle*{2}}
\put(0,0){\circle*{2}}
\put(20,0){\circle*{2}}
\qbezier[25](0, 0)(10, 15)(20, 0)
\qbezier[25](0, 0)(10, -15)(20, 0)\kern 1.9em\ 
}

\def\Figh{\kern 0.25em\
\put(10, 0){\circle*{2}}
\put(10, 7.5){\circle*{2}}
\put(0,0){\circle*{2}}
\put(20,0){\circle*{2}}
\put(0,0){\line(20,0){20}}
\qbezier[25](0, 0)(10, 15)(20, 0)
\qbezier[25](0, 0)(10, -15)(20, 0)\kern 1.9em\  
}

\def\Figi{\kern 0.25em\
\put(6, -6.5){\circle*{2}}
\put(14, -6.5){\circle*{2}}
\put(0,0){\circle*{2}}
\put(20,0){\circle*{2}}
\put(0,0){\line(20,0){20}}
\qbezier[25](0, 0)(10, 15)(20, 0)
\qbezier[25](0, 0)(10, -15)(20, 0)\kern 1.9em\ 
}

\def\Figj{\kern 0.25em\
\put(0,0){\circle*{2}}
\put(20,0){\circle*{2}}
\put(0,0){\line(20,0){20}}
\qbezier[25](0, 0)(10, 15)(20, 0)
\qbezier[25](0, 0)(10, -15)(20, 0)\kern 1.9em\ 
}

\def\Gifa{\kern 0.25em\
\put(5,-9){\thicklines{\line(1, 2){9}}}
\put(0,0){\circle*{2}}
\put(20,0){\circle*{2}}
\put(0,0){\line(20,0){20}}
\qbezier(0, 0)(10, 15)(20, 0)
\qbezier(0, 0)(10, -15)(20, 0)\kern 1.9em\ 
}

\def\Gifb{\kern 0.25em\
\put(4,-8){\thicklines{\line(1, 2){9}}}
\put(10, 0){\circle*{2}}
\put(0,0){\circle*{2}}
\put(20,0){\circle*{2}}
\put(0,0){\line(20,0){20}}
\qbezier[25](0, 0)(10, 15)(20, 0)
\qbezier[25](0, 0)(10, -15)(20, 0)\kern 1.9em\ 
}

\def\Gifc{\kern 0.25em\
\put(5,-9){\thicklines{\line(1, 2){9}}}
\put(10, -7.5){\circle*{2}}
\put(0,0){\circle*{2}}
\put(20,0){\circle*{2}}
\put(0,0){\line(20,0){20}}
\qbezier[25](0, 0)(10, 15)(20, 0)
\qbezier[25](0, 0)(10, -15)(20, 0)\kern 1.9em\ 
}

\def\Gifd{\kern 0.25em\
\put(5,-9 ){\thicklines{\line(1, 2){9 }}}
\put(10, -7.5){\circle*{2}}
\put(0,0){\circle*{2}}
\put(20,0){\circle*{2}}
\put(0,0){\line(20,0){20}}
\qbezier(0, 0)(10, 15)(20, 0)
\qbezier(0, 0)(10, -15)(20, 0)\kern 1.9em\ 
}

\def\Gife{\kern 0.25em\
\put(5,-9){\thicklines{\line(1, 2){9}}}
\put(0,0){\circle*{2}}
\put(20,0){\circle*{2}}
\put(0,0){\line(20,0){20}}
\qbezier[25](0, 0)(10, 15)(20, 0)
\qbezier[25](0, 0)(10, -15)(20, 0)\kern 1.9em\ 
}

\begin{document}
\title{TWO-LOOP CROSSOVER SCALING FUNCTIONS OF THE $O(N)$ MODEL}

\author{DENJOE O'CONNOR$^{1,\ast}$, J.~A.~SANTIAGO$^{2,\dagger}$, 
C.~R.~STEPHENS$^{3,\S}$ and A.~ZAMORA$^{2,\ddagger}$}
\address{
$^1$ School of Theoretical Physics, Dublin Institute for Advanced Studies, 
10 Burlington Road, Dublin 4, Ireland\\
$^2$ Departamento de Matem\'aticas Aplicadas y Sistemas, Universidad 
Aut\'onoma Metropolitana - Cuajimalpa, M\'exico D.F. 01120, Mexico\\
$^3$ Instituto de Ciencias Nucleares, Universidad Nacional 
Aut\'onoma de M\'exico, Apartado Postal 70-543, M\'exico D.F. 04510,
Mexico}
\address{$^{\ast}${denjoe@stp.dias.ie}\\
         $^{\dagger}${jsantiago@correo.cua.uam.mx}\\
         $^{\S}${stephens@nucleares.unam.mx}\\
         $^{\ddagger}${zamora@correo.cua.uam.mx}}

\maketitle

\begin{history}
\received{Day Month Year}
\revised{Day Month Year}
\end{history}

\begin{abstract}
Using Environmentally Friendly Renormalization, we present an analytic 
calculation of the series for the renormalization constants that describe the
equation of state for the $O(N)$ model in the whole critical region.
The solution of the beta-function equation, for the running coupling to order 
two loops, exhibits crossover between the strong coupling fixed point, 
associated with the Goldstone modes, and the Wilson-Fisher fixed point. The 
Wilson functions $\gamma_\lambda$, $\gamma_\varphi$ and $\gamma_{\varphi^2}$, and 
thus the effective critical exponents associated with renormalization of the 
transverse vertex functions, also exhibit non-trivial crossover between these 
fixed points.

\end{abstract}

\ccode{PACS numbers: 64.60.ae, 64.60.De, 64.60.F-}

\section{Introduction}
\label{Intro}

The equation of state for the $O(N)$ model remains a subject of great interest 
(see, for instance, Refs.~\refcite{Zinn-Justin,Pelisseto} for recent reviews).
It exhibits crossover behavior between three distinct asymptotic regimes 
---the critical region approached along the critical isotherm, the critical 
region approached along the critical isochore, and, finally, the coexistence 
curve. For $N=1$, the longitudinal correlation length remains finite away from 
the critical point on the coexistence curve, while for $N>1$, the existence 
of Goldstone bosons leads to infrared singularities.\cite{Lawrie,Horner} 
The problem of encapsulating these distinct scaling behaviors within one 
overall scaling function has been solved through an {\it ab initio} 
derivation from an underlying microscopic model.\cite{EqAnalytic}
Specifically, in that work we obtained the equation of state for 
the $O(N)$ model using only the Landau--Ginzburg--Wilson (LGW) Hamiltonian by 
implementing an Environmentally Friendly Renormalization (EFR) 
Group\cite{Stephens} which tracks the crossover between the fixed points that 
control the different asymptotic regimes.

EFR is a formalism within a general
framework of perturbative renormalization and the renormalization group, 
specifically designed to describe crossover phenomena, where the effective
degrees of freedom at different scales can be quite distinct, leading to
different scaling regimes and associated exponents. To do this, a successful 
renormalization should track the evolving nature of the effective degrees of 
freedom as a function of scale and that as the latter depend on the 
``environment'' the reparametrization chosen should also depend of it. To 
illustrate this, if one considered an interacting field theory in a three 
dimensional box of size $L$, one could renormalize the theory in an $L$ 
independent fashion. When one considered physics on scales $\kappa\sim L^{-1}$ 
one would find that the theory was perturbatively ill defined, whereas an 
appropriate $L$ dependent renormalization made perturbative sense. The reason 
for this, of course, is that the effective degrees of freedom in the system 
are explicitly $L$ dependent. An $L$ independent renormalization ignores
this important physical fact. The only fluctuations being absorbed into
the renormalized parameters in this case are $L$ independent, no matter
what renormalization scale one chooses. $L$ here is the parameter which
induces the crossover and therefore a good renormalization scheme should 
be $L$ dependent. In principle, basically any system will exhibit crossover 
behavior in some regime. Some pertinent examples are: systems with a 
bicritical point\cite{Fisher}, bulk-surface crossovers\cite{Binder}
and dimensional crossover\cite{Barber}.

The advantage of this method relative to standard RG techniques is that it
describes perturbatively the crossovers between any and all fixed points as
opposed to the perturbative regime around one single fixed point. The 
disadvantage is that the Feynman diagrams that enter in the perturbative
calculations are computed in the relevant environment where the finite part
not just the asymptotic divergence is crucial. Applying EFR to study the 
crossovers inherent in the equation of state we obtain explicit functional 
forms that obey all required 
analyticity properties and require no phenomenological input, only the three
Wilson functions $\gamma_\lambda$, $\gamma_\varphi$, and $\gamma_{\varphi^2}$ 
which are deduced from within the theory. In particular, the case $N=1$ was
treated analytically in the one-loop approximation in Ref.~\refcite{EqAnalytic}. 
In a recent study\cite{Eq1Loop} we carried out the task again in the one-loop 
approximation, but now to include all $N\ge 1$.

In this paper we continue the extension towards the two-loop approximation
of the universal equation of state by deriving the Wilson scaling functions to 
this order. The main motivation for doing this is to contribute with the 
crossover functions required to obtain the equation of state, and to calculate 
the first non-trivial correction to the critical effective exponent $\eta$, 
which follows directly from the Wilson function $\gamma_\varphi$. Such a result 
is useful also as a checkout as no analytic form can be accessed to this order
even for $N=1$. As a considerable part of the formidable task of getting the 
equation of state for the $O(N)$ model to order two loops, we consider it 
worthwhile to present the calculation in this publication. To begin with, 
in Section \ref{RG-Wilson-f} we outline the renormalization group representation
for the {\it ab initio} formulation that will be considered throughout this 
paper. Then, in Section \ref{Series} we obtain explicit expressions to the order
of two loops for the bare vertex functions, the renormalization constants and 
the Wilson functions. In Section \ref{Num} we derive the two-loop order Wilson 
functions and solve the beta-function equation of the running coupling, whose 
curve exhibits the crossover between the Wilson-Fisher and strong-coupling fixed 
points for $N>1$. The corresponding crossovers in the Wilson functions, and 
therefore in the effective critical exponents, are also presented in this 
section. Finally, we draw concluding remarks in Section \ref{Conclu}.

\section{Renormalization Group Representation}
\label{RG-Wilson-f}

The model is described by the standard LGW Hamiltonian with $O(N)$ symmetry
\begin{equation}
{\cal H}[{\bf\varphi}]=\int\!
\hbox{d}^{d}x\left({1\over 2}\nabla\varphi^a\nabla\varphi^a
\!+{1\over 2}r(x)\varphi^a\varphi^a
\!+{\lambda_B\over 4!}(\varphi^a\varphi^a)^2\!\right),
\label{hamilt}
\end{equation}
which describes an $N$-component scalar field ${\bf\varphi}$ in a 
$d$-dimensional space.
Here $r$ denotes the bare mass parameter, $r=r_c+t_B$, with $r_c$ being the 
value of $r$ at the critical temperature $T_c$ and 
$t_B=\Lambda^2\!\left(\frac{T-T_c}{T_c}\right)$, where $\Lambda$ is the 
microscopic scale. The value $r_c$ can naturally be interpreted in statistical 
mechanics as $r_c\propto T_c-T_m$, where $T_m$ is the critical temperature 
predicted by mean field theory. As is well known, an additive renormalization 
for $r_c$ is first necessary to compensate for the critical temperature shift; 
then a further multiplicative renormalization of $t_B$ is needed.

The generator of connected correlation functions,
$W$, is given by
\begin{equation}
W[H_a]=\ln Z,
\end{equation}
where $Z$ is the functional integral over the order parameter fields 
$\varphi^a$, with Hamiltonian (\ref{hamilt}) and an external source $H_a(x)$
\begin{equation}
Z[H]=\int[d\varphi]e^{-{\cal H}[\varphi]+\int d^dx H_a\varphi^a}.
\end{equation}
The connected correlation functions are obtained by repeated functional 
differentiation of $W$ with respect to $t_B(x)$ and $H_a(y)$. We denote these by
\begin{equation}
G^{(N,M)}_{a_1\dots a_N}(x_1,\dots, x_N, y_1,\dots, y_M).
\end{equation}
In the same fashion, the vertex functions $\Gamma^{(N, M)}_{a_1\dots a_N}$
are obtained by functional differentiation with respect to $\bar\varphi(x)$ 
and $t(x)$ of the effective action $\Gamma[\bar\varphi]$, which is given as 
the Legendre transform
\begin{equation}
\Gamma[\bar\varphi]=-W[H_a]+\int d^dx\,\, H_a(x)\bar\varphi^a(x),
\end{equation}
where $\bar\varphi^a(x)$ is the physical magnetization of the system defined 
as
\begin{equation}
\bar\varphi^a(x)=Z^{-1}\left.\frac{\delta Z}{\delta H_a(x)}\right|_{H_a=0}.
\end{equation}
In the ordered phase two types of modes exist: those along the external field
$H_a$ and those perpendicular to it. If we denote by $n^a$ to the unit vector 
in the direction of the external field, then by using the projectors
\begin{equation}
P_\ell^{ab}=n^an^b ,\qquad  P_t^{ab}=\delta^{ab}-n^an^b,
\label{projectors}
\end{equation}
we can write a general vertex function as $\Gamma^{(N,M)}_{\ell\dots\ell\,t\dots t}$.
When all subscripts are equal to $\ell$ or $t$, we compact them to one.
For instance, $\Gamma^{(N,M)}_{t\dots t}$ is denoted $\Gamma_t^{(N,M)}$. 
Additionally, if there are no $\varphi^2$ insertions (i.e. $M=0$), the second
superscript is omitted. That is, we write $\Gamma^{(N,0)}=\Gamma^{(N)}$.\\

As a consequence of the Ward identities of this model, all vertex functions 
can be expressed in terms of the transverse vertex functions. For instance, 
from the equation of state $\Gamma^{(1)}_a= H_a$ we have $\Gamma^{(1)}_t=0$ 
and $\Gamma^{(1)}_\ell=H$, so that use of the Ward identity 
$\Gamma^{(1)}_\ell=\Gamma^{(2)}_t\bar\varphi$ yields
\begin{equation}
\Gamma^{(2)}_t\bar\varphi=H,\,\qquad \Gamma^{(1)}_t=0.
\label{eqstate-(2)} 
\end{equation}
Decomposing $\Gamma^{(2)}_{ab}$ produces $\Gamma^{(2)}_\ell$, $\Gamma^{(2)}_t$ 
and $\Gamma^{(2)}_{\ell\,t}$. Ward identities then imply
\begin{equation}
\Gamma^{(2)}_\ell=\Gamma^{(2)}_t +\frac{\Gamma^{(4)}_t}{3}\bar\varphi^2 
\quad{\rm and} \quad \Gamma^{(2)}_{\ell\,t}=0.
\label{Ward-id}
\end{equation}
Analogously, one may express any vertex function in terms of the 
$\Gamma^{(N,M)}_t$. In this sense, the transverse vertex functions are the 
building blocks of the theory. The Wilson functions $\gamma_i$, 
to be defined shortly, can in particular be written in terms of these 
functions.\cite{Stephens}\\

Due to the existence of large fluctuations in the critical regime,
a renormalization of the microscopic bare parameters of the form
\begin{eqnarray}
t(m_t,\kappa)&=&Z_{\varphi^2}^{-1}(\kappa)t_B(m_t), \\
\lambda(\kappa)&=&Z_{\lambda}(\kappa)\lambda_B, \\
\bar\varphi(\kappa)&=&Z_{\varphi}^{-1/2}(\kappa)\bar\varphi_B,
\label{ren-bare-p}
\end{eqnarray}
must be imposed, where $\kappa$ is an arbitrary renormalization scale
and $m_t$ is the inverse transverse correlation length. 
The renormalized parameters satisfy the differential equations
\begin{eqnarray}
\kappa\frac{\hbox{d} t(\kappa)}{\hbox{d}\kappa}&=&
\gamma_{\varphi^2}(\kappa)t(\kappa),\qquad\qquad\,\,\,
\hbox{where}\qquad\quad
\gamma_{\varphi^2}(\kappa)=-\left.\kappa\frac{\hbox{d}}{\hbox{d}\kappa}
\ln Z_{\varphi^2} \right|_{c}, \\
\kappa\frac{\hbox{d} \lambda(\kappa)}{\hbox{d}\kappa}&=&
\gamma_{\lambda}(\kappa)\lambda(\kappa),\qquad\qquad\,\,\,\,
\hbox{where}\qquad\quad
\gamma_{\lambda}(\kappa)=\left.\kappa\frac{\hbox{d}}{\hbox{d}\kappa}
\ln Z_{\lambda} \right|_{c}, \\
\kappa\frac{\hbox{d}\bar\varphi(\kappa)}{\hbox{d}\kappa}&=&-\frac{1}{2}
\gamma_{\varphi}(\kappa)\bar\varphi(\kappa),\qquad\quad\hbox{where}\qquad\quad
\gamma_{\varphi}(\kappa)=\left.\kappa\frac{\hbox{d}}{\hbox{d}\kappa}
\ln Z_{\varphi} \right|_{c},
\label{ren-diff-eq}
\end{eqnarray}
where on the right-hand side are the Wilson functions associated with
this coordinate transformation and the derivatives are taken along an
appropriately chosen curve in the phase diagram, which we here denote
by $c$. In this paper we are interested precisely in finding the
$\gamma_i$ as crossover scaling functions to the order of two loops.

Integration of the RG equation for any multiplicatively renormalizable
$\Gamma_t^{(N,M)}$ yields
\begin{equation}
\Gamma_t^{(N,M)}(t,\lambda,\bar\varphi)=\hbox{e}^{\int_{\kappa}^{m_t}
\left(\frac{N}{2}\gamma_{\varphi}-M\gamma_{\varphi^2}\right)
\frac{{d}x}{x}}
\Gamma_t^{(N,M)}(t(\kappa),\lambda(\kappa),\bar\varphi(\kappa)).
\label{transv-vertex}
\end{equation}
The renormalization constants $Z_\varphi$, $Z_{\varphi^2}$ and $Z_\lambda$
are fixed by imposing the explicitly magnetization-dependent 
normalization conditions on the transverse correlation functions
\begin{eqnarray}
\left.\partial_{p^2}\Gamma_t^{(2)}(p,t(\kappa,\kappa),\lambda(\kappa),
\bar\varphi(\kappa),\kappa)\right|_{p^2=0}&=&1, \\
\Gamma_t^{(2,1)}(0,t(\kappa,\kappa),\lambda(\kappa),
\bar\varphi(\kappa),\kappa)&=&1, \\
\Gamma_t^{(4)}(0,t(\kappa,\kappa),\lambda(\kappa),
\bar\varphi(\kappa),\kappa)&=&\lambda,
\label{norm-cond}
\end{eqnarray}
while the condition
\begin{equation}
\kappa^2=\Gamma_t^{(2)}(0,t(\kappa,\kappa),\lambda(\kappa),
\bar\varphi(\kappa),\kappa)
\label{gauge-fix-cond}
\end{equation}
serves as a gauge fixing condition that relates the sliding renormalization
scale $\kappa$ to the physical temperature $t$ and magnetization $\bar\varphi$.
Physically, $\kappa$ is a fiducial value of the nonlinear scaling field
$m_t$.

Besides $m_t$, the other nonlinear scaling field we use to parametrize
our results is
\begin{equation}
m_{\varphi}^2=\frac{1}{3}
\frac{\Gamma_t^{(4)}\bar\varphi^2}{\left.\partial_{p^2}\Gamma_t^{(2)}\right|_{p^2=0}}
,\label{sc-field-cond}
\end{equation}
which is RG invariant. It represents the anisotropy in the masses of the
longitudinal and transverse modes and is related to the stiffness constant
$\rho_s=\bar\varphi^2\left.\partial_{p^2}\Gamma_t^{(2)}\right|_{p^2=0}$ via
$m_{\varphi}^2=\frac{1}{3}\lambda\rho_s$. With this renormalization 
prescription one may determine the Wilson scaling functions in terms of 
the nonlinear scaling fields $m_t$ and $m_\varphi$, as the transverse and
longitudinal propagators that appear in all perturbative diagrams can
be parametrized in terms of them.

\section{Perturbative Series}
\label{Series}

\subsection{The bare correlation functions}

Within the {\it ab initio} formulation we are using, there appears the difficulty
of calculating Feynman diagrams that are no longer simple numbers but
functions of the variables $m_t$ and $m_\varphi$ instead. In the one-loop 
approximation it is possible to evaluate the integrals involved analytically;
however to higher orders this is no longer possible and this, numerically,
complicates derivation of the equation of state.

In this section we present a perturbative expansion of the correlation 
functions and show explicitly their dependence on the nonlinear scaling 
functions. We start from the two-loop effective action
%\begin{widetext}
\begin{eqnarray}
\Gamma \left[ \overline{\varphi }\right]
&=&\int \hbox{d}^dx {1\over 2}\left[\overline{\varphi}^{\imath}(x)
\left(\Delta+r\right)\overline{\varphi}^{\imath}(x)+\frac{\lambda }{4!}
\overline{\varphi}^4(x)\right] \nonumber\\  
&&+\frac{\lambda }{4!}\int \hbox{d}^dx\Big[
3G_{\ell }^{2}\left( x,x\right) +2\left( N-1\right) G_{\ell }\left(
x,x\right) G_{t}\left( x,x\right) \nonumber\\
&&+\left( N^{2}-1\right) G_{t}^{2}\left(x,x\right) \Big]
-\frac{\lambda ^{2}}{36}\int d^dx\,\,d^dy\,\, \overline{\varphi }^{\imath }\left(
x\right) \Big[ 3G_{\ell }^{3}\left( x,y\right)\nonumber\\
&& + \left( N-1\right) G_{\ell
}\left( x,y\right) G_{t}^{2}\left( x,y\right) \Big] \overline{\varphi }%
^{\imath }\left( y\right) ,
\label{gammatwo}
\end{eqnarray}
% \end{widetext}
which embodies the physics of the system. Notice that we are implicitly working
with bare quantities and to simplify the writing we will be using diagrammatic
notation. We represent the longitudinal propagator by a solid line and the 
transverse propagator by a dotted line: 
$G_{\ell}^{-1}=p^2+r+{\lambda\over 2}{\overline{\varphi}}^2$ and 
$G_{t}^{-1}=p^2+r+{\lambda\over 6}{\overline{\varphi}}^2$ respectively.
It can be observed that in the symmetric ordered phase (magnetization 
$\overline\varphi=0$) the two propagators are equivalent. Moreover, for models
in the Ising universality class ($N=1$) the terms mixing propagators have no 
contribution in the correlation function.

To obtain the two point correlation function, we take two functional 
derivatives of $\Gamma[\overline{\varphi}]$ respect to the order parameter 
$\overline\varphi$. After a large but otherwise direct calculation, we find 
the two-loop approximation for the two point correlation function in momenta 
space
%\begin{widetext}
\begin{eqnarray}
&&\Gamma _{\imath\jmath}^{\left( 2\right) }\left( p\right)
=-\frac{\lambda }{4!}\bigg[6\Diga\Digb
+\frac{2}{3}\left( N-1\right)  \Diga\Figb\nonumber\\
&& +2\left( N-1\right) \Digb\Figa + \frac{2}{3}\left( N^{2}-1\right)
\Figa\Figb\bigg]\delta _{\imath\jmath}  \nonumber \\
&&+\frac{\lambda ^{3}}{4!}\bigg[ 12\Diga\Digc +6\Digb^{2}+\frac{4}{3}
\left( N-1\right)\Figb\Digb \nonumber\\ 
&&+\frac{4}{9}(N-1)\Diga\Figg+4\left( N-1\right)
\Figa\Digc \nonumber \\
&&+\frac{2}{9}\left( N^{2}-1\right)\Figb^{2}+\frac{4}{9}\left(
N^{2}-1\right) \Figa\Figg\bigg]\overline{\varphi }^{\imath}
\overline{\varphi }^{\jmath }\nonumber \\
&&-\frac{\lambda ^{2}}{36}\bigg[ 6\Digg\delta _{\imath\jmath}-18\lambda
\left(\Digf+\Digf_{0}\right)\overline{\varphi }^{\imath}
\overline{\varphi }^{\jmath}-9\lambda \overline{\varphi }^{2}
\Digf\delta _{\imath\jmath}\nonumber\\
&&+9\lambda ^{2}\overline{\varphi }^{2}\left( \Digh +\Digh_{0}\right) 
\overline{\varphi }^{\imath}\overline{\varphi }^{\jmath }  
+9\lambda ^{2}\overline{\varphi }^{2}\bigg(\Dige \nonumber\\
&&+\Dige_{0}\bigg) \overline{%
\varphi }^{\imath}\overline{\varphi }^{\jmath }\bigg] 
-\left( N-1\right) \frac{\lambda ^{2}}{36}
\bigg[2\Figj\lambda_{\imath\jmath}-2\lambda
\bigg( \Fige \nonumber\\
&&+\Fige_{0}\bigg) 
\overline{\varphi }^{\imath}\overline{\varphi }%
^{\jmath }-\frac{2}{3}\lambda\left(\Figd+\Figd_{0}\right) \overline{\varphi }%
^{\imath}\overline{\varphi }^{\jmath }  \nonumber \\
&&-\lambda \overline{\varphi }^{2}\Fige\delta _{\imath\jmath}+\lambda ^{2}
\overline{\varphi }^{2}\left(\Figc+\Figc_{0}\right) \overline{\varphi }%
^{\imath}\overline{\varphi }^{\jmath }\nonumber\\
&&+\frac{1}{3}\lambda ^{2}\overline{\varphi }^{2}\left(
\Figh +\Figh_{0}\right) \overline{\varphi }^{\imath}\overline{\varphi }^{\jmath } 
\nonumber \\
&&-\frac{2}{3}\lambda \overline{\varphi }^{2}\Figd
\delta _{\imath\jmath}-\frac{2}{3}%
\lambda \left(\Figd +\Figd_{0}\right) \overline{\varphi }^{\imath}
\overline{\varphi }^{\jmath }  \nonumber \\
&&+\frac{1}{3}\lambda^2\overline{\varphi }^{2}\bigg(\Figh 
+\Figh_{0}+\frac{1}{3}\left(\Figf +\Figf_{0}\right)\nonumber\\
&& +\frac{2}{3}\left(\Figi +\Figi_{0}\right)\bigg) 
\overline{\varphi }^{\imath}\overline{\varphi }^{\jmath }\bigg] ,
\label{gamma2p}
\end{eqnarray}
%\end{widetext}
where $p$ denotes the external momenta flowing through the diagrams, the index
$0$ indicating evaluation of the corresponding diagram at the point $p^2=0$. 
By using the projectors (\ref{projectors})
onto this expression one may identify both the 
transverse $\Gamma_{t}^{\left( 2\right) }\left( p\right)$ and the longitudinal 
$\Gamma _{\ell}^{\left( 2\right) }\left( p\right)$ two point correlation 
functions. The transverse two point vertex function is given by      
%\begin{widetext}
\begin{eqnarray}
&&\Gamma _{t}^{\left( 2\right) }\left( 0\right) =t+\frac{\lambda }{6}
\overline{\varphi }^{2}+\frac{\lambda }{2}\Diga +\left( N-1\right) 
\frac{\lambda }{6}\Figa  \nonumber \\
&&-\frac{\lambda }{4!}\bigg[ 6\Diga\Digb+\frac{2}{3}\left( N-1\right)\Diga
\Figb+2\left( N-1\right)\Digb\Figa\nonumber\\
&&+\frac{2}{3}\left( N^{2}-1\right)\Figa\Figb
\bigg]  -\frac{\lambda ^{2}}{36}\left[ 6\Digg\ -
9\lambda \overline{\varphi }^{2}\Digf\right]\nonumber\\
&&-\left( N-1\right) \frac{\lambda ^{2}}{36}\left[ 2\Figj-\lambda 
\overline{\varphi }^{2}\left(\Fige +\frac{2}{3}\Figd\right)\right] ,
\label{Gammat2p}
\end{eqnarray}
%\end{widetext}
where we have added the terms arising from the one-loop contribution. The
derivative of this function respect to the external momenta $p$ is also
required. For this we find
%\begin{widetext}
\begin{eqnarray}
\partial _{p^{2}}\Gamma _{t}^{\left( 2\right) }\left( 0\right)&=&
1-\frac{\lambda ^{2}}{36}\left[ 6\Gifa\ -9\lambda \overline{\varphi }^{2}
\Gifd \right]\nonumber\\ 
&&-\left( N-1\right) \frac{\lambda ^{2}}{36}\left[ 2\Gife -\lambda 
\overline{\varphi }^{2}\left( \Gifb +\frac{2}{3}\Gifc\right) \right] .
\label{PartialGammat2}
\end{eqnarray}
%\end{widetext}
To obtain this result we have taken the derivative and then evaluated at the 
point $p^2=0$. We emphasize this fact with a diagonal line crossing the 
diagrams.

The four point correlation function can be calculated by taking four 
derivatives in Eq. (\ref{gammatwo}) or alternatively by using the Ward 
identity (\ref{Ward-id}). By using the latter we get to
%\begin{widetext}
\begin{eqnarray}
&&\Gamma _{t}^{\left( 4\right) }\left( 0\right) =\lambda -\frac{3}{2}
\lambda^{2}\left(\Digb +\frac{N-1}{9}\Figb\right)  
+\frac{\lambda ^{3}}{8}\bigg[12\Diga\Digc \nonumber\\
&&+6\Digb^{2}+\frac{4}{3}
\left( N-1\right)\Figb\Digb 
+\frac{4}{9}(N-1)\Diga\Figg\nonumber\\  
&&+4\left( N-1\right) \Figa\Digc+\frac{2}{9}\left( N^{2}-1\right) 
\Figb^{2}\nonumber\\
&&+\frac{4}{9}\left( N^{2}-1\right)\Figa\Figg \bigg]\nonumber\\
&&+\frac{3}{2}\lambda
^{3}\left[ 2\Digf -\lambda \overline{\varphi }^{2}\left(
\Digh +\Dige\right) \right] \nonumber\\ 
&&+\left( N-1\right) \frac{\lambda ^{3}}{12}\bigg[4\Fige +\frac{8}{3}
\Figd \nonumber\\
&&-2\lambda \overline{\varphi }^{2}
\left( \frac{2}{3}\Figh +\frac{1}{9}
\Figf +\Figc +\frac{2}{9}\Figi\right) \bigg] .
\label{Gammat4}
\end{eqnarray}
%\end{widetext}
To complete this stage of the calculation we need the correlation function 
$\Gamma_{t}^{\left( 2,1\right) }$, which can be obtained if one derivative of the
function $\Gamma_t^2(p)$ respect to the parameter $t$ is taken. The resulting
expression is
%\begin{widetext}
\begin{eqnarray}
&&\Gamma _{t}^{\left( 2,1\right) } =1-\frac{\lambda }{2}\Digb -%
\frac{N-1}{6}\lambda\Figb 
+\frac{\lambda ^{2}}{4!}\bigg[6\bigg( 2\Diga\Digc\nonumber\\
&& +\Digb^{2}\bigg) 
+4\left( N-1\right) \left( \frac{1}{3}\Diga\Figg +\Figa\Digc\right)
\nonumber\\
&&+\frac{8}{3}\left( N-1\right) \Digb\Figb \nonumber\\ 
&&+\frac{2}{3}\left( N^{2}-1\right) \left( 2\Figa\Figg+\Figb^{2}\right)\bigg]
\nonumber\\
&&+\frac{\lambda ^{2}}{4}\left[ 2\Digf-2\lambda \overline{%
\varphi }^{2}\left(\Digh +\Dige \right) \right]  \nonumber \\
&&+\left( N-1\right) \frac{\lambda ^{2}}{18}\bigg[\Figd + \Fige +\Figd -\lambda 
\overline{\varphi }^{2}\left(\Figh +\Figc \right)\nonumber\\
&&-\frac{1}{3}\lambda \overline{\varphi }^{2}\left( \Figf + \Figh
+2\Figi\right) \bigg] .
\label{Gammat21}
\end{eqnarray}
%\end{widetext}

\subsection{Wave function renormalization}

We now move on to consider the wave function renormalization. To proceed,
we shall use the relationships that define both scaling variables 
$m_{\varphi }^{2}$ and $m_{t}^{2}$ in the one-loop approximation, and then 
perform an inversion of these at the same order. The next step is to replace 
the resulting expressions into the one-loop terms within the correlation 
functions and then make the expansion up to two loops. By performing these 
expansions into the first two terms of Eq. (\ref{Gammat4}) we get to the 
four point correlation function
%\begin{widetext} 
\begin{eqnarray}
&&\Gamma _{t}^{\left( 4\right) }\left( 0\right) =\lambda _{B}-\frac{3}{2}
\lambda _{B}^{2}\left( \Digb +\frac{N-1}{9}\Figb \right)
+\frac{\lambda _{B}^{3}}{8}\bigg[ 6\Digb^{2}\nonumber\\
&&+\frac{4}{3}\left(N-1\right)\Figb\Digb +\frac{2}{9}\left( N^{2}-1\right)
\Figb^{2}\bigg] +\frac{3}{2}\lambda _{B}^{3}\bigg[ 2\Digf \nonumber\\
&& -\lambda _{B}\overline{\varphi }_{B}^{2}\left(\Digh + \Dige \right) \bigg] 
+\left( N-1\right)\frac{\lambda _{B}^{3}}{12}\bigg[ 4\Fige  +\frac{8}{3}
\Figd\nonumber\\
&&-2\lambda _{B}\overline{\varphi }_{B}^{2}\left( \frac{2}{3}\Figh 
+\frac{1}{9}\Figf +\Figc +\frac{2}{9}\Figi\right) \bigg]\nonumber\\
&&+\frac{\lambda _{B}^{3}}{9}\left( N-1\right) \left[ \Figa-\Diga\right]\Figg 
+\frac{3}{2}\lambda _{B}^{4}\overline{\varphi }_{B}^{2}\bigg[ \Digb\nonumber\\
&& +\frac{N-1}{9}\Figb\bigg] \Digc \,.
\label{Gammat4r}
\end{eqnarray}
%\end{widetext}
By repeating the same procedure into the one-loop terms of 
$\Gamma_{t}^{\left( 2,1\right) }$, one obtains
%\begin{widetext}
\begin{eqnarray}
&&\Gamma _{t}^{\left( 2,1\right) }\left( 0\right) =1-\frac{\lambda _{B}}{2}
\left(\Digb +\frac{N-1}{3}\Figb\right)+\frac{\lambda _{B}^{2}}{4!}
\bigg[ 6\Digb^{2}\nonumber\\ 
&&+\frac{8}{3}\left( N-1\right) \Figb\Digb+\frac{2}{3}\left(
N^{2}-1\right)\Figb^2\bigg]  \nonumber \\
&&+\frac{\lambda _{B}^{2}}{4}\left[ 2\Digf-2\lambda _{B}\overline{
\varphi }_{B}^{2}\left(\Digh +\Dige \right) \right]  
+\left( N-1\right) \frac{\lambda _{B}^{2}}{18}\bigg[ 2\Figd \nonumber\\
&&+ \Fige -\lambda _{B}\overline{\varphi }_{B}^{2}\left(\Figh +\Figc \right)
-\frac{1}{3}\lambda _{B}\overline{\varphi }_{B}^{2}\bigg(\Figf +\Figh\nonumber\\
&& +2\Figi \bigg) \bigg]+\frac{\lambda _{B}^{2}}{9}\left( N-1\right) 
\left[ \Figa -\Diga \right]\Figg\nonumber\\
&& +\frac{1}{2}\lambda _{B}^{3}\overline{\varphi }_{B}^{2}\left[\Digb 
+\frac{N-1}{9}\Figb \right] \Digc \,.
\label{Gammat21r}
\end{eqnarray}
%\end{widetext}
Using these expressions we shall calculate the renormalization constants that 
are necessary to find the Wilson functions to the order of two loops.

\subsection{Renormalization constants}

In this section we calculate the renormalization constants or $Z$ 
functions. These are defined in terms of the bare correlation functions
that we found in the previous section. For the function $Z_\lambda$ we can 
write
%\begin{widetext}
\begin{eqnarray}
&&Z_{\lambda } =1-\frac{3}{2}\lambda _{B}^{2}\left( \Digb +\frac{N-1}{9}
\Figb\right)+\frac{\lambda _{B}^{2}}{8}\bigg[ 6\Digb^{2}\nonumber\\
&&+\frac{4}{3}\left(N-1\right)\Figb\Digb+\frac{2}{9}\left( N^{2}-1\right)
\Figb^{2}\bigg] +\frac{3}{2}\lambda _{B}^{2}\bigg[ 2\Digf \nonumber\\
&&-\lambda _{B}\overline{\varphi }_{B}^{2}\left(\Digh +\Dige \right) \bigg]
+\left( N-1\right) \frac{\lambda _{B}^{2}}{12}\bigg[\Fige +\frac{8}{3}
\Figd\nonumber\\
&&-2\lambda _{B}\overline{\varphi }_{B}^{2}\left( \frac{2}{3}
\Figh +\frac{1}{9}\Figf +\Figc +\frac{2}{9}\Figi\right) \bigg] \nonumber \\
&&+\frac{\lambda _{B}^{2}}{18}\left[ 6\Gifa -9\lambda _{B}
\overline{\varphi }_{B}^{2}\Gifd \right]+\left( N-1\right) 
\frac{\lambda _{B}^{2}}{18}\bigg[ 2\Gife\nonumber\\
&& -\lambda _{B}\overline{\varphi }_{B}^{2}\left( \Gifb +\frac{2}{3}\Gifc 
\right) \bigg] +\frac{\lambda _{B}^{2}}{9}\left( N-1\right) \left[ \Figa 
-\Diga \right]\Figg \nonumber\\
&&+\frac{3}{2}\lambda _{B}^{3}\overline{\varphi }_{B}^{2}\left[ \Digb 
+\frac{N-1}{9}\Figb\right]\Digc \,,
\label{zetalambda}
\end{eqnarray}
%\end{widetext}
whereas for $Z_{\varphi }^{-1}$ the expression is
%\begin{widetext}
\begin{eqnarray}
Z_{\varphi }^{-1} &=& 1-\frac{\lambda _{B}^{2}}{36}\left[ 6\Gifa -9\lambda _{B}
\overline{\varphi }_{B}^{2}\Gifd\right]-\left( N-1\right) 
\frac{\lambda _{B}^{2}}{36}\bigg[ 2\Gife\nonumber\\
 &&-\lambda _{B}\overline{\varphi }_{B}^{2}\left( \Gifb +\frac{2}{3}\Gifc  
\right) \bigg] ,
\label{zetaphi}
\end{eqnarray}
%\end{widetext}
and similarly for $Z_{\varphi^{2}}^{-1}$ one finds 
%\begin{widetext}
\begin{eqnarray}
&&Z_{\varphi ^{2}}^{-1} =-\frac{\lambda _{B}}{2}\left( \Digb +\frac{N-1}{3}
\Figb \right) +\frac{\lambda _{B}^{2}}{4!}\bigg[ 6\Digb^{2}\nonumber\\
&&+\frac{8}{3}\left( N-1\right)\Figb\Digb +\frac{2}{3}\left( N^{2}-1\right) 
\Figb^{2}\bigg]  \nonumber\\
&&+\frac{\lambda _{B}^{2}}{4}\left[ 2\Digf -2\lambda _{B}\overline{%
\varphi }_{B}^{2}\left(\Digh + \Dige \right) \right]  \nonumber \\
&&+\left( N-1\right) \frac{\lambda _{B}^{2}}{18}\bigg[2\Figd + \Fige -%
\lambda _{B}\overline{\varphi }_{B}^{2}\left(\Figh + \Figc\right) \nonumber\\
&&-\frac{1}{3}\lambda _{B}\overline{\varphi }_{B}^{2}\left(\Figf + \Figh
+2\Figi \right) \bigg]  \nonumber \\
&&+\frac{\lambda _{B}^{2}}{36}\left[ 6\Gifa -9\lambda _{B}
\overline{\varphi }_{B}^{2}\Gifd \right]+\left( N-1\right) 
\frac{\lambda _{B}^{2}}{36}\bigg[ 2\Gife\nonumber\\
&&-\lambda _{B} \overline{\varphi }_{B}^{2}\left( \Gifb +\frac{2}{3}\Gifc
\right) \bigg] +\frac{\lambda _{B}^{2}}{9}\left( N-1\right) \left[ \Figa 
-\Diga \right] \Figg \nonumber\\
&&+\frac{1}{2}\lambda _{B}^{3}\overline{\varphi }_{B}^{2}\left[ \Digb 
+\frac{N-1}{9}\Figb\right] \Digc \,.
\label{zetaphi2}
\end{eqnarray}
%\end{widetext}
Notice that we have been working with the bare coupling $\lambda_B$, which
we now write explicitly in these expressions.

\subsection{The Wilson functions}

The Wilson functions are defined in terms of the renormalization constants
$Z_\lambda$, $Z_\varphi$ and $Z_{\varphi^2}$ that we found in the previous section. 
From the definition of $\gamma_\lambda$, by using Eq. (\ref{zetalambda}), one 
obtains
%\begin{widetext}
\begin{eqnarray}
&&\gamma _{\lambda } =-\frac{3}{2}\lambda D_{\kappa }\left( \Digb +\frac{N-1}{9}
\Figb\right) -\lambda ^{2}D_{\kappa }\bigg[ \frac{3}{2}\Digb^{2}\nonumber\\
&&+ \frac{1}{3}\left( N-1\right) \Digb\Figb -\frac{1}{18}\left( N-1\right) 
\Figb^{2}\bigg]\nonumber\\ 
&&+\frac{3}{2}\lambda ^{2}\left[ 2D_{\kappa }\Digf -\lambda \overline{%
\varphi }^{2}D_{\kappa }\left(\Digh +\Dige \right) \right] \nonumber \\
&&+\left( N-1\right) \frac{\lambda ^{2}}{12}\bigg[ 4D_{\kappa }\Fige 
+\frac{8}{3}D_{\kappa }\Figd\nonumber\\ 
&&-\frac{2}{3}\lambda \overline{\varphi }^{2}D_{\kappa
}\left( 2\Figh +\frac{1}{3}\Figf +3\Figc\right)\bigg]\nonumber\\
&&+\frac{\lambda ^{2}}{18}\left[ 6D_{\kappa }\Gifa -9\lambda 
\overline{\varphi }^{2}D_{\kappa }\Gifd\right]\nonumber\\ 
&&+\left(N-1\right) \frac{\lambda ^{2}}{18}\left[ 2D_{\kappa }\Gife -\lambda
\overline{\varphi }^{2}D_{\kappa }\left( \Gifb +\frac{2}{3}\Gifc\right) \right]
\nonumber\\
&&+\frac{3}{2}\lambda^{3}\overline{\varphi }^{2}D_{\kappa}\left[ \Digb 
+\frac{N-1}{9}\Figb\right] \Digc \nonumber\\
&&+ \frac{\lambda^{2}}{9}\left( N-1\right)\left[ D_{\kappa}\left( \Figa -\Diga 
\right) \Figg-\frac{\lambda\overline\varphi^2}{3}D_{\kappa}\Figi \right] ,
\label{gammalambdar}
\end{eqnarray}
%\end{widetext}
where we have used ${d/d\ln k}:=D_{\kappa }$ to simplify the writing.

Similarly, from the definition of $\gamma_\varphi$ and using Eq. (\ref{zetaphi}),
we get
%\begin{widetext}
\begin{eqnarray}
\gamma _{\varphi } &=& \frac{\lambda ^{2}}{36}\left[ 6 D_{\kappa }\Gifa 
-9\lambda \overline{\varphi }^{2}D_{\kappa}\Gifd \right]\nonumber\\
&&+\left( N-1\right) \frac{\lambda ^{2}}{36}\left[ 2 D_{\kappa}\Gife -\lambda 
\overline{\varphi }^{2}D_{\kappa }\left(\Gifb +\frac{2}{3}\Gifc\right)\right] ,
\label{gammaphi}
\end{eqnarray}
%\end{widetext}
and finally from Eq. (\ref{zetaphi2}) we obtain for the Wilson function 
$\gamma_{\varphi^2}$, 
%\begin{widetext}
\begin{eqnarray}
&&\gamma _{\varphi ^{2}} =-\frac{\lambda }{2}D_{\kappa }\left( \Digb 
+\frac{N-1}{3}\Figb \right)  \nonumber \\
&&-\lambda ^{2}\bigg[ \frac{1}{4}D_{\kappa }\Digb^{2}+\frac{1}{18}
\left( N-1\right) \Figb D_{\kappa }\Digb\nonumber\\
&& +\frac{2}{9}\left( N-1\right) \Digb D_{\kappa }\Figb
-\frac{1}{18}\left( N-1\right) D_{\kappa }\Figb^{2}\bigg]\nonumber\\
&& +\frac{\lambda ^{2}}{2}\left[ D_{\kappa }\Digf -\lambda\overline{\varphi}^{2}
D_{\kappa }\left(\Digh +\Dige \right) \right]  \nonumber \\
&&+\left( N-1\right) \frac{\lambda ^{2}}{18}\bigg[ D_{\kappa }\left(
2\Figd + \Fige \right) -\lambda \overline{\varphi }^{2}D_{\kappa }\left(
\Figh +\Figc\right)\nonumber\\
&&-\frac{1}{3}\lambda \overline{\varphi }^{2}D_{\kappa}\left(\Figf + \Figh 
\right) \bigg]+\frac{\lambda ^{2}}{36}\left[ 6 D_{\kappa }\Gifa\ -9\lambda 
\overline{\varphi }^{2}D_{\kappa }\Gifd\right]\nonumber\\
&&+\left(N-1\right) \frac{\lambda ^{2}}{36}\left[ 2 D_{\kappa }\Gife 
-\lambda \overline{\varphi }^{2}D_{\kappa }\left( \Gifb +\frac{2}{3}\Gifc 
\right) \right]\nonumber\\
&&+\frac{1}{2}\lambda^{3}\overline{\varphi }^{2}D_{\kappa}\left[ \Digb 
+\frac{N-1}{9}\Figb \right]\Digc \nonumber\\
&&+ \frac{\lambda^{2}}{9}\left( N-1\right)\left[ D_{\kappa}
\left( \Digb -\Figb\right)\Figg -\frac{\lambda\overline\varphi^2}{3}D_{\kappa}
\Figi\right] .
\label{gammaphi2}
\end{eqnarray}
%\end{widetext}
Notice that we have applied the $D_{\kappa}$ to the renormalization constants
and then replaced $\lambda_B$ in terms of the renormalized coupling $\lambda$.
Also note that in these expressions all the diagrams are given in terms of the
nonlinear scaling fields $m_t$ and $m_\varphi$ explicitly.

\section{3D Two-loop Results }
\label{Num}

As anticipated in Section \ref{Intro}, we are interested in obtaining
expressions for the $\gamma_i$ as crossover scaling functions. The natural 
variable is $z$, and so the next stage consists in writing the Wilson 
functions in terms of the nonlinear scaling fields $z=m_t/m_\varphi$. For 
instance, the Wilson function $\gamma_{\lambda}$ which is cubic in the coupling 
(see Eq. (\ref{gammalambdar})) turns out a quadratic in terms of $z$. It is
no difficult to express the Feynman diagrams appearing in the Wilson functions
in terms of $z$. By defining the functions
%\begin{widetext}
\begin{eqnarray}
f_1(z)
&=& D_{\kappa }\left( \Digb +\frac{N-1}{9}\Figb\right) , \nonumber\\
&\bbuildrel{=}_{}^{3d}& -\frac{1}{8\pi ^{2}}\left( \frac{N-1}{9}+\left( 1+
\frac{1}{z^{2}}\right) ^{-3/2}\right) , \nonumber\\
f_2(z)
&=&-D_{\kappa }\bigg[ \frac{3}{2}\Digb^{2}+ \frac{1}{3}\left( N-1\right)
\Digb\Figb -\frac{1}{18}\left( N-1\right) \Figb^{2}\bigg] , \nonumber\\
&\bbuildrel{=}_{}^{3d}& \frac{1}{64\pi ^{2}}\left( -\frac{N-1}{9}+\frac{
2\left( N-1\right) }{3}\frac{1+\frac{1}{2z^{2}}}{\left( 1+\frac{1}{z^{2}}
\right) ^{3/2}}+3\left( 1+\frac{1}{z^{2}}\right) ^{-2}\right) , \nonumber \\
f_3(z)
&=&\left[ 2D_{\kappa }\Digf -\lambda \overline{\varphi }^{2}
D_{\kappa }\left(\Digh +\Dige \right) \right] , \nonumber \\
&\bbuildrel{=}_{}^{3d}&-\frac{1}{16\pi ^{3}}\left( \frac{2\pi }{3}\left( 1+
\frac{1}{z^{2}}\right) ^{-2}-\frac{3}{z^{2}}\left( 1+\frac{1}{z^{2}}\right)^{-3}
\left( \frac{\pi }{9}+{1.4\over 2}\right) \right) , \nonumber \\
f_4(z)
&=&\left[ 4D_{\kappa }\Fige +\frac{8}{3}
D_{\kappa }\Figd -\frac{2}{3}\lambda \overline{\varphi }^{2}D_{\kappa}
\left( 2\Figh +\frac{1}{3}\Figf +3\Figc\right)\right] , \nonumber\\
f_5(z)
&=&\left[ 6D_{\kappa }\Gifa -9\lambda 
\overline{\varphi }^{2}D_{\kappa }\Gifd\right] , \nonumber\\ 
&\bbuildrel{=}_{}^{3d}&\frac{1}{72\pi ^{2}}\left( 1+\frac{1}{z^{2}}\right)^{-2}
-\frac{1}{24\pi ^{2}}\frac{1}{z^{2}}\left( 1+\frac{1}{z^{2}}\right)^{-3} ,
\nonumber \\
f_6(z)
&=&\left[ 2D_{\kappa }\Gife-\lambda \overline{\varphi }^{2}D_{\kappa }
\left( \Gifb +\frac{2}{3}\Gifc\right) \right] , \nonumber\\ 
f_7(z)
&=&\lambda^{3}\overline{\varphi }^{2}D_{\kappa}\left[ \Digb +\frac{N-1}{9}
\Figb\right] \Digc \,,  \nonumber\\
&\bbuildrel{=}_{}^{3d}& -\frac{3}{256\pi ^{2}}\frac{1}{z^2}\bigg( 4\left( 1
+\frac{1}{z^{2}}\right) ^{-3}+\frac{\left( N-1\right) }{9}\left( 1
+\frac{1}{z^{2}}\right)^{-3/2} \nonumber\\
&&+\frac{\left( N-1\right) }{3}\left( 1+\frac{1}{z^{2}}\right)^{-5/2}\bigg) ,
\nonumber\\
f_8(z)
&=&\frac{1}{9}\left[ D_{\kappa}\left( \Figa -\Diga \right) 
\Figg-\frac{\lambda\overline\varphi^2}{3}D_{\kappa}\Figi \right] ,\nonumber
\end{eqnarray}
%\end{widetext}
the  crossover scaling functions $\gamma_\lambda$ and $\gamma_\varphi$ can be 
written respectively as
\begin{eqnarray}
\gamma_{\lambda }\left( z\right)&=&-\frac{3}{2}f_{1}\overline{\lambda }
+\bigg( f_{2}+\frac{3}{2}f_{3}+\frac{N-1}{12}f_{4}+\frac{1}{18}f_{5}
+\frac{N-1}{18}f_{6} \nonumber\\
&&+\frac{3}{2}f_{7}+\left( N-1\right) f_{8}\bigg) 
\overline{\lambda}^{2} , \\
\label{gammlam}
\gamma_{\varphi}\left( z\right)&=&\left(\frac{1}{36}f_5 
+\frac{N-1}{36}f_6\right)\overline{\lambda }^{2} .
\end{eqnarray}
Further, definition of the functions
\begin{eqnarray}
g_1(z)
&=& D_{\kappa }\left( \Digb +\frac{N-1}{3}\Figb\right) , \nonumber\\
&\bbuildrel{=}_{}^{3d}& -\frac{1}{8\pi ^{2}}\left( \frac{N-1}{3}+\left( 1+
\frac{1}{z^{2}}\right) ^{-3/2}\right) , \nonumber\\
g_2(z)&=& \bigg[ \frac{1}{4}D_{\kappa }\Digb^{2}+\frac{1}{18}
\left( N-1\right) \Figb D_{\kappa }\Digb\nonumber\\
&& +\frac{2}{9}\left( N-1\right) \Digb D_{\kappa }\Figb
-\frac{1}{18}\left( N-1\right) D_{\kappa }\Figb^{2}\bigg] , \nonumber\\
&\bbuildrel{=}_{}^{3d}& -{\Gamma(1/2)^2\over (4\pi )^3}\bigg[ {1\over 2} 
\left(1+{1\over z^2}\right)^{-2}+{N-1\over 18}\left(1+{1\over z^2}\right)^{-3/2}
\nonumber\\
&&+{2(N-1)\over 9} \left(1+{1\over z^2}\right)^{-1/2}+{N-1\over 9}\bigg] ,
\nonumber\\
g_3(z)&=&\left[ D_{\kappa }\Digf -\lambda \overline{\varphi }^{2}D_{\kappa }
\left(\Digh +\Dige \right) \right] , \nonumber \\
&\bbuildrel{=}_{}^{3d}&-{1\over (4\pi )^3}\bigg[ {4\pi\over 3} 
\left(1+{1\over z^2}
\right)^{-2} -{12\over z^2}\left(1+{1\over z^2}\right)^{-3}\left({\pi\over 9}
+{1.4\over 2}\right)\bigg] , \nonumber\\
g_4(z)&=& \bigg[ D_{\kappa }\left(
2\Figd + \Fige \right) -\lambda \overline{\varphi }^{2}D_{\kappa }\left(
\Figh +\Figc\right)\nonumber\\
&&-\frac{1}{3}\lambda \overline{\varphi }^{2}D_{\kappa}\left(\Figf + \Figh 
\right) \bigg] , \nonumber\\
g_5(z)&=& \left[ D_{\kappa}
\left( \Digb -\Figb\right)\Figg -\frac{\lambda\overline\varphi^2}{3}D_{\kappa}
\Figi\right] , 
\end{eqnarray}
allows one to write $\gamma_{\varphi^2}$ in the form
\begin{eqnarray}
\gamma_{\varphi^2}\left( z\right) &=&\frac{1}{2}f_7-\frac{1}{2}g_1
\overline{\lambda}+\bigg(-g_2+\frac{1}{2}g_3+\frac{N-1}{18}g_4+\frac{1}{36}f_5
\nonumber\\
&& +\frac{N-1}{36}f_6+\frac{N-1}{9}g_5\bigg)\overline{\lambda}^{2} .
\end{eqnarray}
Note that all the crossover scaling Wilson functions are finite in the limits 
of small and large values of $z$. For the functions $f_{i}\left( z\right)$ and 
$g_{i}\left( z\right)$ for which it is not possible to obtain an exact 
analytical expression, we have verified using Mathematica that their behavior 
is finite in both asymptotic limits.

From the definition of $\gamma_{\lambda}$ and its relation to the $\beta$ 
function, we get the $\beta$-function equation for the dimensionless coupling
$\overline{\lambda}$:
\begin{equation}
\kappa \frac{d\overline{\lambda }}{d\kappa }=-(4-d)\overline{\lambda }
+\gamma _{\lambda }\overline{\lambda} .
\label{blam}
\end{equation}
As discussed in Ref.~\refcite{Stephens}, one attempts to reconstruct the Wilson 
functions from their series to a given order in the loop expansion using some 
method of resummation. This is necessary as the direct perturbative series 
arising from the EFR scheme are divergent asymptotic series for any coupling 
strength, so 
that physical results can only be accessed from resummation procedures.
Pad\'e resummation is one technique, based on rational functions having the same 
power series expansion as the original series to the given order, that has been 
successfully used and is widely accepted for resumming perturbative series since 
the early works of Baker,\cite{Baker} and so we use it here to obtain a resummed 
series for the beta function; i.e. right-hand side of Eq. (\ref{blam}). 
A noteworthy fact is that using different Pad\'e approximants one may estimate 
errors in the resummed series. This is especially useful for higher loop 
computations.

There also exist more sophisticated techniques that can be used for resummation 
of 
asymptotic series, such as the Pad\'e-Borel and conformal mapping methods. The 
former applies the Pad\'e approximation to the Borel transform, whereas the latter 
is an improvement to the Pad\'e-Borel method based on the mapping of a complex 
plane into the unit disk. More efficient techniques are based on re-expansions
of the asymptotic truncated series in terms of special basis functions which 
are chosen to possess precisely the analytic behavior responsible for the 
divergence of the original series. These, and the method of variational 
perturbation theory, which is a systematic extension of a variational 
approximation to path integrals, require information on the behavior of the 
series, so they are suitable for higher order expansions (see Ref.~\refcite{Kleinert}
for a review of these methods).

\begin{figure}[ht]
\begin{center}
\framebox{
\includegraphics[width=2.7in,angle=0]{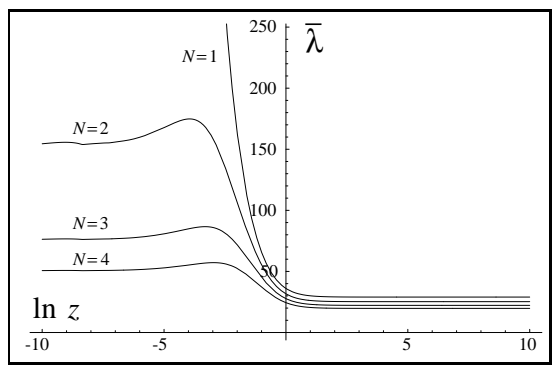}}\quad
\end{center}
\caption{\small The Pad\'e-resummed coupling $\overline\lambda (z)$, 
for several values of $N$. Note the crossover to the strong 
fixed point for  $z\rightarrow 0$.}
\label{lambda}
\end{figure}

For our two-loop calculations, we use for simplicity the Pad\'e approximant. 
That is, we solve numerically the $\left[ 2/1 \right]$ Pad\'e-resummed 
differential equation arising from Eq. (\ref{blam}). The solutions we find are 
shown in Fig.~\ref{lambda}.
There, one may see the coupling parameter interpolating continually between the 
fixed points of the model. The presence of the Wilson-Fisher fixed point 
$\overline{\lambda }^{\ast}$ can be observed for large numerical values of $z$ 
and, for $N>1$, the fixed point associated with the coexistence curve in the 
limit $z\rightarrow 0$. Notice the relative decrease of the maximum, on the left
of the curve, as $N$ increases. More precise values in the asymptotic regimes, 
obtained from higher order approximations that introduce external numerical data,
are known in the literature. This leads one to expect the same qualitative 
behavior of the curve to higher orders in the loop approximation. For 
completeness, in 
Table~\ref{T1} we present the asymptotic values for the coupling in the limit
$z\rightarrow 0$ and in Table~\ref{T2} the values corresponding to the 
Wilson-Fisher fixed point. 
\begin{table}[ph]
\tbl{Asymptotic values of the 3D coupling and Wilson functions in the 
limit $z\rightarrow 0$.}
{\begin{tabular}{@{}ccccc@{}}
\toprule
$N$  &  ${\overline\lambda}$  &  $\gamma_\lambda$  &  $\gamma_\varphi$  & 
$\gamma_{\varphi^2}$ \\
\colrule 
1 &           $\infty$  & 0 & 0 & 0 \\ 
2 &            154.4    & 1 & 0 & 1 \\ 
3 & \hphantom{0}76.6    & 1 & 0 & 1 \\ 
4 & \hphantom{0}50.6    & 1 & 0 & 1 \\
\botrule
\end{tabular} 
\label{T1}}
\end{table}
\begin{table}[ph]
\tbl{Asymptotic values of the 3D coupling, Wilson 
functions and critical exponents at the WF critical point.}
{\begin{tabular}{@{}ccccccccc@{}}
\toprule
$N$  &  ${\overline\lambda}^*$  &  $\gamma_\lambda^*$  &  $\gamma_\varphi^*$  & 
$\gamma_{\varphi^2}^*$ & $\beta$ & $\delta$ & $\nu$ & $\gamma$  \\
\colrule
1 & 29.01 & 1 & 0.033 & 0.43 & 0.33 & 4.8 & 0.64 & 1.26 \\ 
2 & 25.26 & 1 & 0.033 & 0.52 & 0.35 & 4.8 & 0.68 & 1.33 \\ 
3 & 22.23 & 1 & 0.032 & 0.59 & 0.37 & 4.8 & 0.71 & 1.39 \\ 
4 & 19.76 & 1 & 0.031 & 0.64 & 0.38 & 4.8 & 0.73 & 1.45 \\
\botrule
\end{tabular}
\label{T2}}
\end{table}
However, as these values come from a two-loop calculation, no greater
precision is expected. What is worthwhile noticing is the fact that, within
our {\it ab initio} calculation and without external input data, we capture the
crossover-function character of the running coupling interpolating from one 
fixed point to the other.
By substituting the numerical solution for the coupling into the 
Pad\'e-resummed $\gamma_{\lambda }\left( z\right)$ function, we obtain the 
behavior of the Wilson functions showing the continuum crossover between the 
Wilson-Fisher fixed point and the fixed point associated with the coexistence 
curve. 

\begin{figure}[ht]
\begin{center}
\framebox{
\includegraphics[width=2.7in,angle=0]{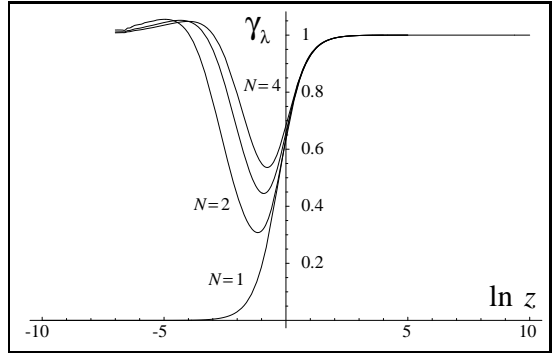}}
\end{center}
\caption{\small The  Wilson function $\gamma_\lambda (z)$ showing that
$\gamma_{\lambda}\rightarrow 1$ in the limits $z\rightarrow\infty$ and
$z\rightarrow 0$, except for $N=1$.}
\label{gammalambda}
\end{figure}

\begin{figure}[ht]
\begin{center}
\framebox{
\includegraphics[width=2.7in,angle=0]{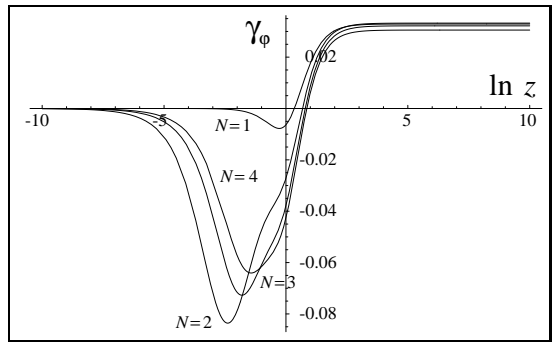}}
\end{center}
\caption{\small The Wilson function $\gamma_\varphi (z)$.}
\label{gammaphifig}
\end{figure}

\begin{figure}[ht]
\begin{center}
\framebox{
\includegraphics[width=2.7in,angle=0]{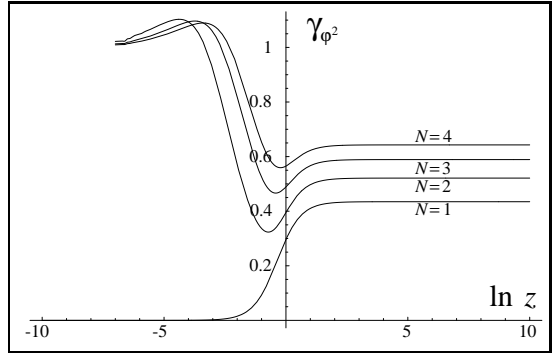}}
\end{center}
\caption{\small The Wilson function $\gamma_{\varphi^2}(z)$. Notice
the mean field behavior of the Ising model, $N=1$, in the limit 
$z\rightarrow 0$ whereas $\gamma_{\varphi ^{2}}\rightarrow 1$ for $N>1$.}
\label{gammaphi2fig}
\end{figure}

The crossover scaling function $\gamma_{\lambda}(z)$ in 
Fig.~\ref{gammalambda} provides information on the effective dimension of the 
system\cite{Stephens}. Given that the dimension employed to evaluate the 
expressions is $d=3$, asymptotically the value is $1$, as expected.
Nevertheless, it is of theoretical and experimental interest the local 
minimum values of this function for $N>1$. For $N=1$, the limit
$\gamma_{\lambda}\rightarrow 0$ shows mean field behavior as the fluctuations
are suppressed. On the other hand, the crossover scaling function 
$\gamma_{\varphi}(z)$ directly corresponds to the effective exponent $\eta$, 
see Fig.~\ref{gammaphifig}. As one would expect, in the Wilson-Fisher fixed
point, this takes values that approximate those known from higher order
calculations. We stress once again that, in the two-loop order approximation 
that we consider in this work, the contribution comes from the {\it ab initio}
calculation and no attempt has been made at improving numerical values 
using external data. The numerical values that we obtain are, nevertheless, 
provided in table~\ref{T2}. Once more, in the limit $z\rightarrow 0$ the 
fluctuations are suppressed and this can be observed in the zero value of 
this anomalous exponent.
Finally, the crossover scaling function $\gamma_{\varphi^2}$ in 
Fig.~\ref{gammaphi2fig} also shows non-trivial crossover. In the asymptotic 
limit $z\rightarrow \infty$, this function is related to the critical exponent 
$\nu$. It can be observed that the curves $\gamma_{\varphi^2}(z)$, for $N>1$, 
tend to follow the trajectory described by the Ising model in the limit 
$z\rightarrow 0$, but in the end they separate from it. Again in this case,
just as a reference, the asymptotic values of this function and for completeness
the values of the critical exponents have also been included in both tables.

\section{Conclusions}
\label{Conclu}

By using EFR it is possible, from an {\it ab initio} calculation, to derive 
a parametric form for the equation of state of the $O(N)$ model that has all 
desired analiticity properties in the three distinct asymptotic regimes.
To order one loop it is even possible to obtain an analytic expression for
$N=1$ in dimension three. We see that the fundamental building blocks for
calculations are the Wilson functions. However, despite the fact that they
play a privileged role they enter into expressions for physical quantities
in a non-trivial way, such as in integrals. In the case where the Wilson
functions must be computed numerically in the first place this involves
numerical subtleties that are not present in standard calculations, such as 
perturbation expansions in $\varepsilon$ or $1/N$ for problems where, for 
instance, the upper critical dimension or the symmetry of the order parameter 
change, respectively. In reality this is to be expected, calculating a 
crossover scaling function is much more complicated than calculating an 
exponent.

From the EFR formalism it would seem that once the Wilson functions are
calculated at a given order, the derivation of the equation of state at the 
same order should be somehow straightforward. However, the Feynman diagrams 
appearing in the Wilson functions to order two loops are crossover functions 
themselves and the difficulty precisely resides in systematically sum diagrams
whose divergences do cancel. This is a problem we are currently sorting out. 
In this paper we have performed explicit two-loop order calculations of the 
transverse correlation functions, the renormalization constants and the Wilson 
functions. By solving numerically the beta-function equation, we have captured 
the crossover between the critical fixed point and the fixed point associated 
with the coexistence curve. The Wilson functions also show the crossover. In 
the limit $z\to\infty$, the Wilson-Fisher fixed point is approached and 
$\gamma_i\to\gamma_i^{WF}$ with $\gamma_\lambda=1$ for $d=3$. In contrast, in the 
limit $z\to 0$ the strong-coupling fixed point is approached and 
$\gamma_i\to\gamma_i^{SC}$. For $N>1$ the Goldstone bosons dominate and 
$\gamma_\lambda=\gamma_{\varphi^2}=1$. For $N=1$ however, this fixed point is mean 
field like as fluctuations are suppressed and $\gamma_i\to 0$.

\section*{Acknowledgments}

JAS would like to thank Roc\'\i o Mondrag\'on and Claudio Santiago for support 
and motivation during the several stages of this work.

\appendix

\section{Derivatives of Feynman Integrals}

We summarize here expressions for some of the diagrams appearing in the 
numerical functions $f_{i}\left( z\right)$ and $g_{i}\left( z\right)$.
In terms of the functions
\begin{eqnarray}
f\left( x,y\right) &=&x\left( 1-x\right) \left( 1-y\right) +y ,
\label{ap1} \nonumber\\
g(x,y,z) &=& f\left( x,y\right) +y\left( 1-x\right) z^{-2} , \nonumber
\end{eqnarray}
we can write
%\begin{widetext}
\begin{eqnarray}
D_{\kappa }\Fige &=&A_{d}\int_{0}^{1}\int_{0}^{1}dx\, dy\frac{x^{1-d/2}\left(
y\left( 1-x\right) \right) ^{2-d/2}f(x,y)}{g\left( x,y,z\right) ^{5-d}} ,
\nonumber \label{g1}\\
D_{\kappa }\Figd &=&A_{d}\int_{0}^{1}\int_{0}^{1}dx\, dy\frac{\left( x\left(
1-x\right) \right) ^{2-d/2}\left( 1-y\right) y^{1-d/2}f(x,y)}{g\left(
x,y,z\right) ^{5-d}} ,\nonumber
\label{g2}
\end{eqnarray}
%\end{widetext}
where $A_{d}=-\frac{2\left( 4-d\right) \Gamma \left( 4-d\right) }{\left(
4\pi \right) ^{d}}\frac{1}{\kappa ^{2\left( 4-d\right) }}$. Analogously, 
we have
%\begin{widetext}
\begin{eqnarray}
D_{\kappa }\Figh&=&B_{d}\int_{0}^{1}\int_{0}^{1}dx\, dy\frac{y^{3-d/2}\left(
x\left( 1-x\right) \right) ^{2-d/2}f(x,y)}{g\left( x,y,z\right) ^{6-d}} ,
\nonumber \label{g3} \\
D_{\kappa}\Figf &=&B_{d}\int_{0}^{1}\int_{0}^{1}dx\, dy\frac{\left( y\left(
1-x\right) \right) ^{2-d/2}x^{3-d/2}\left( 1-y\right) f(x,y)}{g\left(
x,y,z\right) ^{6-d}} , \nonumber \\
D_{\kappa }\Figc &=&\frac{B_{d}}{2}\int_{0}^{1}\int_{0}^{1}dx\, dy\frac{\left(y
\left( 1-x\right) \right) ^{3-d/2}x^{1-d/2}f(x,y)}{g\left( x,y,z\right)^{6-d}} ,
\nonumber 
\end{eqnarray}
%\end{widetext}
where $B_{d}=-\frac{2\left( 5-d\right) \Gamma \left( 5-d\right) }{\left(
4\pi \right) ^{d}}\frac{1}{\kappa ^{2\left( 5-d\right) }}$. Finally for the
diagrams with the diagonal line crossing them, we find
%\begin{widetext}
\begin{eqnarray}
D_{\kappa }\Gife &=&-A_{d}\int_{0}^{1}\int_{0}^{1}dx\, dy\frac{
y^{2-d/2}\left( 1-y\right) \left( x\left( 1-x\right) \right) ^{2-d/2}f(x,y)}
{g\left( x,y,z\right) ^{5-d}} , \nonumber \label{h1} \\
D_{\kappa }\Gifb &=&-B_{d}\int_{0}^{1}\int_{0}^{1}dx\, dy\frac{
x^{1-d/2}\left( y\left( 1-x\right) \right) ^{3-d/2}\left( 1-y\right) f(x,y)}
{g\left( x,y,z\right) ^{6-d}} , \nonumber \label{h2} \\
D_{\kappa }\Gifc &=&-B_{d}\int_{0}^{1}\int_{0}^{1}dx\, dy\frac{
\left( x\left( 1-x\right) \right) ^{3-d/2}y^{2-d/2}\left( 1-y\right)^{2}
f(x,y)}{g\left( x,y,z\right) ^{6-d}} . \nonumber \label{h3}
\end{eqnarray}
%\end{widetext}

\end{document}